\documentclass[useAMS,usenatbib]{mn2e} 
\usepackage{graphicx}
\usepackage{amsmath}

\title[The Nature of Star Formation in Local Clusters]{ACCESS III: The Nature of Star Formation in the Shapley Supercluster}

\author[Haines et al.]{C. P. Haines,$^{1}$ G. Busarello,$^2$ P. Merluzzi,$^2$ R. J. Smith,$^{3}$ S. Raychaudhury,$^{1}$ \and A. Mercurio,$^2$ G. P. Smith$^{1}$\\
$^{1}$School of Physics and Astronomy, University of Birmingham, Edgbaston, Birmingham, B15 2TT, UK; cph@star.sr.bham.ac.uk\\
$^{2}$INAF - Osservatorio Astronomico di Capodimonte, via Moiariello 16, I-80131 Napoli, Italy\\
$^{3}$Department of Physics, University of Durham, Durham DH1 3LE, UK\\
}

\begin{document}

\maketitle
\label{firstpage}

\begin{abstract}
We present a joint analysis of panoramic {\em Spitzer}/MIPS mid- and far-infrared (MIR/FIR) and {\em GALEX} ultraviolet imaging of the most massive and dynamically active structure in the local Universe, the Shapley supercluster at $z{=}0.048$, covering the five clusters which make up the supercluster core. Combining this with existing spectroscopic data from 814 confirmed supercluster members and 1.4\,GHz radio continuum maps, this represents the largest complete census of star-formation (both obscured and unobscured) in local cluster galaxies to date, extending down to ${\rm SFRs}{\sim}0.$02--0.$05\,{\rm M}_{\odot}{\rm yr}^{-1}$. We take advantage of this comprehensive panchromatic dataset to perform a detailed analysis of the {\em nature} of star formation in cluster galaxies of the kind previously limited to local field galaxy surveys such as SINGS, using several quite independent diagnostics of the quantity and intensity of star formation to develop a coherent view of the types of star formation within cluster galaxies. 

We observe a robust bimodality in the infrared ($f_{24{\mu}{\rm m}}/f_{K}$) galaxy colours, which we are able to identify as another manifestation of the broad split into star-forming spiral and passive elliptical galaxy populations seen in UV-optical surveys.
This diagnostic also allows the identification of galaxies in the process of having their star formation quenched as the infrared analogue to the ultraviolet ``green valley'' population.
The bulk of supercluster galaxies on the star-forming sequence have specific-SFRs consistent with local field specific-SFR--$\mathcal{M}$ relations of Elbaz et al. (2007), and form a tight FIR--radio correlation (0.24\,dex) confirming that their FIR emission is due to star formation. 
We show that 85 per cent of the global SFR is quiescent star formation within spiral disks, as manifest by the observed sequence in the $(L_{IR}/F_{FUV}$)--($FUV{-}NUV$) plane -- the IRX-$\beta$ relation -- being significantly offset from the starburst relation of Kong et al. (2004), while their FIR--radio colours indicate dust heated by low-intensity star formation. 
Just 15 per cent of the global SFR is due to nuclear starbursts. The vast majority of star formation seen in cluster galaxies comes from normal infalling spirals who have yet to be affected by the cluster environment.

\end{abstract}


\section{Introduction}
\label{intro}

\setcounter{footnote}{3}

It is well known that the environment in which a galaxy inhabits has a profound impact upon its evolution in terms of defining both its structural properties and star-formation histories, with passively-evolving ellipticals, S0s and, at lower masses dwarf ellipticals (dEs), dominating cluster cores whereas in field regions galaxies are typically star-forming spirals. These trends have been quantified through the morphology--density \citep{dressler80} and star formation (SF)--density relations \citep[e.g.][]{dressler85,poggianti99,lewis,gomez}. While some of the diminution in star-formation seen in cluster galaxies with respect to the field population could be attributed to the different morphological compositions of the two environments, even at fixed morphology, the star-formation rate (SFR) of cluster galaxies is found to be lower than in the field \citep{balogh00}, indicating separate processes at least partially shape these relations. 

Empirically the S0s of local clusters are mostly replaced by star-forming spirals at $z{\sim}0.5$ \citep[e.g.][]{dressler97,treu} and may be completely absent by $z{\sim}1$ \citep{smith05,postman,desai}. Similarly, we see a marked decline in the cluster dE population out to $z{\sim}1$ \citep{tanaka,delucia}. A simple interpretation is that clusters accreted gas-rich, star-forming spirals at $z{\simeq}0$.5--1 and then these galaxies were transformed somehow into the passive S0s and dEs found in clusters at $z{\simeq}0$. However, it is vital at this point that we place clusters within the current $\Lambda$CDM cosmological models that have proved able to successfully reproduce statistically the growth of cosmic large-scale structure. In these models structure formation occurs hierarchically \citep{springel}, and so clusters are not isolated, but represent the most massive and dynamically immature systems, located at the intersections of filaments, continuously accreting galaxies and galaxy groups along these filaments or directly from the field. Indeed, as \citet{berrier} show, ${\sim}5$0\% of galaxies in local clusters were accreted since $z{\sim}0.4$, meaning that their transformation must be much more wide-spread, rapid, and still ongoing.

Numerous mechanisms have been proposed to deplete the reservoir of gas in infalling spirals, cutting off the fuel supply for further star formation, as well as cause their morphological transformation, resulting in passive S0s and dEs, such as ram-pressure and/or tidal stripping, galaxy merging (including AGN feedback), harassment and starvation \citep[for reviews see e.g.][]{boselli,haines07}. These environmental processes are predicted to quench the star formation over diverse time-scales (e.g. $10^{7-8}$\,yr for ram-pressure stripping; ${\ga}10^{9}$\,yr for starvation). Some mechanisms are additionally expected to include an intermediate starburst and possible subsequent post-starburst phase (e.g. ram-pressure or merger-induced nuclear starbursts). 

A number of studies have attempted to distinguish among the various environmental processes by identifying and classifying various promising classes of transition galaxies, such as post-starburst and/or E+A galaxies \citep{poggianti99}, interaction-induced starbursts \citep{moss,fadda}, and passive/red spirals \citep{vandenbergh,bamford}, which seem to indicate that more than one mechanism is required. However, a critical handicap facing many of these studies is their reliance on a single indicator of current/recent star formation, which may well be sensitive towards only a certain class (or subset) of transition population. For example, many analyses of the impact of the cluster environment were based on spectra obtained from fibre-based spectrographs. These are largely insensitive to the ongoing effects of ram-pressure stripping, which has been observed to produce {\em truncated} H$\alpha$ and H{\sc i} disks in infalling spirals \citep{vogt,crowl,rose}, but which will appear unaffected when observed through fibres covering only the galaxy cores.
Furthermore, early mid-infrared ({\em ISO}) and radio studies of clusters revealed populations of luminous infrared galaxies ($L_{IR}{>}10^{11}L_{\odot}$; LIRGs) and near-LIRGs \citep{duc}, that implied that the SFRs derived from UV or optical diagnostics (e.g. O{\sc ii}) underestimated the true SFRs by ${\sim}1$0--3$0{\times}$ \citep[for a review see][]{metcalfe}. Moreover, the age-dependence of this dust obscuration (the stars responsible for the O{\sc ii} emission are younger and more obscured than those which produce absorption in H$\delta$) can lead to dusty star-bursts being in fact classified as post-starburst galaxies \citep{poggianti00}. 

To fully understand the impact of the cluster environment on galaxy evolution given the above issues, and to distinguish among the diverse physical mechanisms proposed to transform cluster galaxies, a new multi-faceted approach is required.
Firstly, a complete census of star formation in and around clusters is mandatory, based on {\em global} SFR measures, and which fully accounts for dust obscuration. This necessitates a multi-wavelength approach including infrared data sensitive to the obscured SF component. 

Secondly, by comparison of the specific-SFRs (or equivalently birthrate parameter) to the observed tight mass-dependent sequence seen in field spiral galaxies \citep[e.g.][]{elbaz,bothwell}, it becomes possible to distinguish among normal star-forming galaxies, starbursts, passive galaxies, and transitional galaxies (e.g. green-valley objects) whose specific-SFRs are intermediate between the passive and star-forming sequences, and presumably are in the process of having their star formation quenched. Related to this is the ability to distinguish between normally star-forming spirals who are forming stars quiescently over their entire disks and galaxies undergoing nuclear starbursts, i.e. the {\em nature} of the star formation. Even if the measured specific-SFRs are the same, whereas the former is likely evolving quiescently, the latter is likely a temporary state, possibly induced by interaction with its environment, and could thus represent one phase of the transformation of that galaxy. Clearly this can be achieved by resolving where the star formation occurs within a galaxy, but even if all we have is the globally-measured spectral energy distribution (SED), there are several diagnostics within the SED, such as the IRX-$\beta$ relation \citep[e.g.][]{meurer,kong,calzetti05} or the FIR--radio colours \citep{dh02}, sensitive to the intensity of the current star formation within a galaxy.

Finally, to fully distinguish among the diverse environmental processes transforming galaxies, it will often be necessary to resolve its impact on the galaxy, either in terms of its dynamical state (via measurements of its kinematics) or its stellar population (e.g. via colour gradients, H$\alpha$ profiles). For example, ram-pressure stripping quenches star-formation from the outside-in, producing truncated H$\alpha$ profiles, while starvation results in a global decline in the star formation \citep[e.g.][]{boselli06,crowl}. Neither of these processes should affect the kinematics of the galaxy. Instead gravitational mechanisms such as merging, harassment or tidal forces should be recognisable by their kinematic imprints.

\subsection{The ACCESS survey}

It is with all of the above in mind that we are undertaking the ACCESS\footnote{{\em http://www.na.astro.it/ACCESS}} multi-wavelength survey of the Shapley supercluster core (SSC), the most massive and dynamically active structure in the local ($z{<}0.1$) universe. The target has been chosen since the most dramatic effects of the hierarchical assembly of structure on galaxy evolution should occur in supercluster. Here the infall and encounter velocities of galaxies are greatest (${>}1$\,000\,km\,s$^{-1}$), groups and clusters are still merging, and significant numbers of galaxies will be encountering the dense intra-cluster medium (ICM) of the cluster environment for the first time. 

The Shapley supercluster was identified as the largest overdensity of galaxies and clusters in the $z{<}0.1$ Universe in an APM survey of galaxies covering the entire southern sky \citep{sr89} and confirmed by early X-ray studies \citep{sr91,day91}.  In the original Abell catalogue \citep{aco89}, the core of the Shapley supercluster was found to be represented by a single cluster (A3558$\equiv$Shapley~8) of optical richness~4, inconsistent with the observed velocity dispersion of the cluster. Early X-ray observations from {\em Einstein} and {\em ROSAT} \citep{breen94a} resolved this by discovering that the core in fact comprises of a chain three merging clusters, A3558, SC1327-312 and SC1329-313. The SSC in this paper refers to these three clusters, along with the two associated Abell clusters on either side, A3562 and A3556 \citep[see][for a general description]{sos1}.

In \citet[Paper I]{merluzzi} we introduced the survey as well as presented an analysis of the $K$-band data. In \citet[hereafter Paper II]{paper2} we presented the {\em GALEX} ultraviolet and {\em Spitzer} infrared observations of the SSC, producing both ultraviolet and infrared luminosity functions. These data were used to build a {\em complete} and {\em unbiased} census of star formation in galaxies belonging to the SSC which we use here. 
By assembling a comprehensive panchromatic (FUV--FIR) dataset for the SSC we were able to fully account for both obscured and unobscured star-formation, and by dealing with {\em global} photometric measurements, were independent of aperture biases. Using this approach, we showed that just 20 per cent of the global SFR of $327\,{\rm M}_{\odot}{\rm yr}^{-1}$ was visible directly in the ultraviolet continuum, while ${\sim}8$0 per cent was reprocessed by dust and emitted in the infrared. 

In this paper we take advantage of this already comprehensive panchromatic dataset and census of star formation, to which we add 1.4\,GHz radio continuum data from \citet{miller}, in order to examine in unprecedented detail the {\em nature} of star formation in cluster galaxies. Using several quite independent diagnostics of the quantity and intensity of star formation, we carefully develop a coherent picture of the types of star formation activity in these galaxies, measuring the relative importance of quiescent star-formation occurring within normal spiral arms and nuclear star-bursts which may have been triggered by interaction between the galaxy and its environment. In particular, we attempt to discern whether star-formation within cluster galaxies can be considered in a manner consistent with being drawn from the field population, or if there is a significant cluster-specific mode of star-formation.

In {\S}~\ref{sec:data} we present our multi-wavelength datasets, and in {\S}~\ref{sec:bimodal} we combine our {\em Spitzer}/MIPS 24$\mu$m and $K$-band data to establish the bimodal distribution in infrared colour ($f_{24}/f_{K}$) of supercluster galaxies, which can be understood as another manifestation of the well known broad division of galaxies into star-forming spiral and passively-evolving early-type populations. Given the close correspondences between 24$\mu$m and $K$-band luminosities to SFRs and stellar masses respectively, the inference of $f_{24}/f_{K}$ as a natural proxy for specific-SFR is clear. 
In the next two sections we present two apparently disparate and unrelated analyses, the IRX-$\beta$ ({\S}~\ref{sec:irx}) and FIR--radio ({\S}~\ref{sec:radio}) relations. These relations provide two independent diagnostics of the {\em nature} of star formation among the supercluster galaxy population, allowing us to distinguish between quiescent star formation over a spiral disk and intense nuclear starbursts and/or AGN. 
Finally, in {\S}~\ref{sec:discuss} we piece together all of the lines of evidence put forward in Sections~\ref{sec:bimodal}--\ref{sec:radio} to form a coherent picture of the dominant modes of star formation taking place within the Shapley supercluster core, and the diverse populations of transforming galaxies, suggesting a variety of evolutionary pathways from infalling spiral to passive lenticular.

This work is carried out in the framework of the joint research programme ACCESS\footnote{{\it A Complete Census of Star-formation and nuclear activity in the Shapley supercluster}, PI: P. Merluzzi, a collaboration among the Universities of Durham and Birmingham (UK), the Italian National Institute of Astrophysics' Osservatorio di Capodimonte and the Australian National University (ANU).} \citep{merluzzi} aimed at distinguishing among the mechanisms which drive galaxy evolution across different ranges of mass by their interactions with the environment. This requirement to cover a wide mass range encompassing both dwarf and giant populations is particularly relevant given the observed quite different SF--density relations for giant and dwarf galaxies in and around rich clusters \citep[e.g.][]{haines06,haines07}.

When necessary we assume \mbox{$\Omega_{M}{=}0.3$}, \mbox{$\Omega_{\Lambda}{=}0.7$} and \mbox{H$_{0}{=}70\,$km\,s$^{-1}$Mpc$^{-1}$}, such that at the distance of the Shapley supercluster 1\,arcsec is equivalent to 0.96\,kpc.
All magnitudes are quoted in the Vega system unless otherwise stated.

\section{Data}
\label{sec:data}

The new mid-infrared and ultraviolet data analysed in this paper are complemented by existing panoramic optical $B$- and $R$-band imaging from the Shapley Optical Survey \citep[SOS:][]{sos1,sos2} and near-infrared $K$-band imaging \citep{merluzzi}, which covers the same region. 
The ultraviolet and mid-infrared (24$\mu$m, 70$\mu$m) photometry cover essentially the same regions, the bulk of which are also covered by the $K$-band imaging, while the earlier SOS covers a slightly narrower region. This panchromatic dataset covers a common survey area of $\sim$2.2\,deg$^{2}$. The relative coverages of the different aspects of our multi-wavelength survey and the Shapley supercluster core are shown in Figure 1 of Paper II. 

The high quality of the optical images (FWHMs${\sim}0$.7--1.0\,arcsec) has allowed morphological classifications and structural parameters to be derived for all cluster galaxies to ${\sim}M_{R}^{*}{+}3$ \citep{gargiulo} within the automated {\sc 2dphot} environment \citep{labarbera}, while stellar masses were estimated by fitting the observed $BRK$ colours to the model composite stellar populations of \citet{maraston}. 
Full details of the observations, data reduction, and the production of the galaxy catalogues are described in Papers I,II and \citet{sos1}. The characteristics of the photometric surveys are summarized in Table 1.

\begin{table}
\begin{tabular}{cccc} \hline
Wave band & Instrument & Completeness & Ref. \\
($\lambda_{eff}$) &        &  limit  &    \\ \hline
FUV (0.15\,$\mu$m)  & GALEX       & 22.5 (AB)   & [1,2]   \\
NUV (0.23\,$\mu$m) & GALEX      &  22.0 (AB)  & [1,2]   \\
B     (0.44\,$\mu$m)   & ESO2.2m-WFI    & 22.5   &    [3]\\
R    (0.64\,$\mu$m)   & ESO2.2m-WFI    & 22.0   &    [3]\\
K  (2.2\,$\mu$m)     &  UKIRT-WFCAM    & 18.0   &    [4]\\
MIR (23.7\,$\mu$m)   &  {\em Spitzer}/MIPS &  0.35 mJy   &    [2]\\
FIR  (71.4\,$\mu$m)   &  {\em Spitzer}/MIPS &   25 mJy &    [2]\\ \hline
\end{tabular}
\caption{
{\it Column} 1. Waveband.
{\it Column} 2. Instrument.
{\it Column} 3. Magnitude or flux completeness limit for each survey. These limits correspond to the 90\% completeness for GALEX and Spitzer data, 50\% completeness for UKIRT data in the high density environments and 80\% in the low density environments. 100\% completeness for WFI B and R data.
{\it Column } 4. Bibliographic reference for the data: [1] Rawle et al 2008; [2] Paper II; [3] Mercurio et al 2006; [4] Paper I.
}
\end{table}

The Shapley supercluster core has been observed by a variety of redshift surveys \citep{quintana,bardelli98,bardelli00,kaldare,drinkwater,smith07,cava,6df} resulting in 964 galaxies in our $K$-band catalogue having redshift information, of which 814 lie in the velocity range of the SSC ($10\,700{<}cz{<}17\,000$\,km\,s$^{-1}$). Of these, 415 are from the AAOmega-based spectroscopic survey of \citet{smith07}.
We have redshifts for all 107 $K{<}12.3$ ($K{<}K^{*}{+}0.6$) galaxies within $K$-band survey. The 90\% and 50\% spectroscopic completeness limits are $K{=}13.25$ ($K^{*}{+}1.5$) and $K{=}14.3$ ($K^{*}{+}2.5$) respectively.

The panoramic {\em Spitzer} mid-infrared observations of the Shapley supercluster core were carried out over 27--30 August 2008 within the Cycle 5 GO programme 50510 (PI: C.P. Haines). The observations consist of five contiguous mosaics observed with MIPS \citep{rieke} in medium scan mode, providing homogeneous coverage at both 24$\mu$m and 70$\mu$m.

The 24$\mu$m survey limit of 0.35\,mJy corresponds to $L_{IR}{=}7.5{\times}10^{8}L_{\odot}$, while the 70$\mu$m survey limit of 25\,mJy corresponds to $L_{IR}{=}5.7{\times}10^{9}L_{\odot}$ based on comparison to the luminosity-dependend infrared SEDs of \citet{rieke09}.
In total we associate 4663 $f_{24}{>}350{\mu}$Jy sources with galaxies in the $K$-band image, of which 490 have confirmed redshifts, 394 of which place them in the Shapley supercluster. We have redshifts for all $f_{24}{>}26$mJy sources, and 50\% redshift completeness at ${\sim}5$\,mJy. 
At 70\,microns we associate 381 $f_{70}{>}25$mJy sources with galaxies in our $K$-band imaging, of which 194 have confirmed redshifts, 160 of which place them in the Shapley supercluster. 
For more details of the reduction and catalogue production see Paper II.

For the radio, we take the published 1.4\,GHz source catalogues of \citet{miller}, based on his 7\,deg$^{2}$ Very Large Array (VLA) radio continuum survey which encompasses all of our UV--FIR datasets, and reaches $5{\sigma}$ sensitivity of 300$\mu$Jy at a resolution of 16\,arcsec. We thus have full multi-wavelength coverage of the main filamentary structure connecting A3562 and A3558, and the extension to A3556.  

\begin{figure*}
\centerline{\includegraphics[width=170mm]{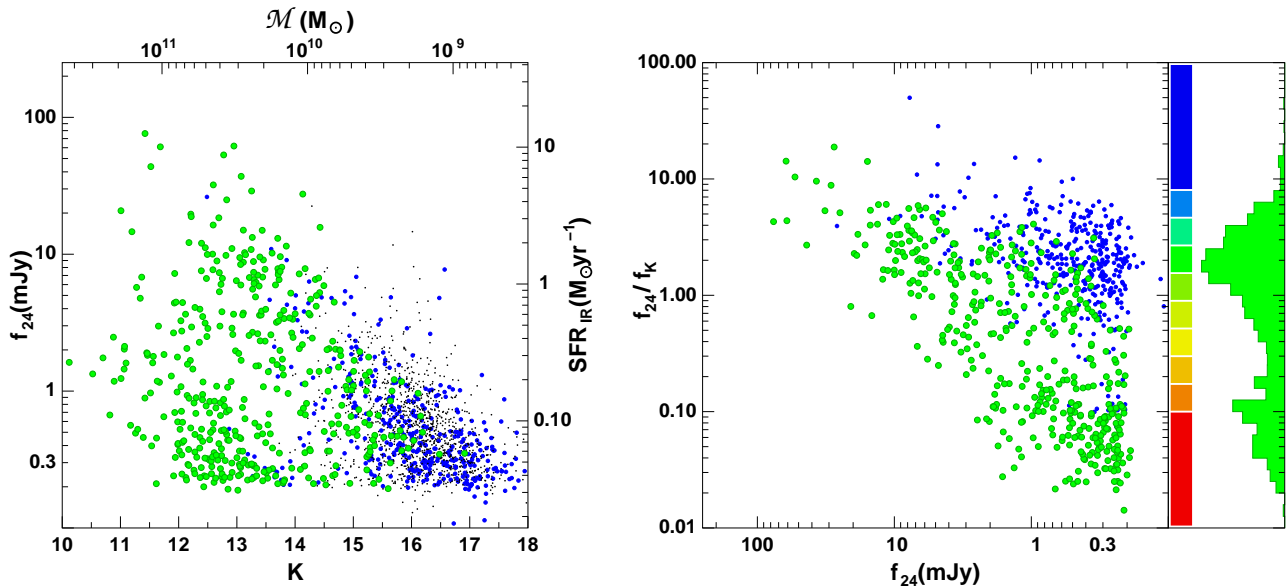}}
\caption{({\em left:}) Observed 24$\mu$m fluxes versus $K$-band magnitude. Large green circles indicate spectroscopically-confirmed supercluster members. Blue symbols indicate those photometrically-selected as supercluster members according to their UV-optical colours, while black dots indicate sources photometrically identified as background galaxies. ({\em right:}) The infrared colour $f_{24}/f_{K}$ versus $f_{24}$ for Shapley supercluster galaxies. The histogram shows the distribution of $f_{24}/f_{K}$ colours of the 24$\mu$m detected supercluster galaxies, demonstrating the clear bimodality in the infrared colours of supercluster galaxies. The vertical band indicates the colour scheme used in the remainder of the article to distinguish galaxies according to their  $f_{24}/f_{K}$ flux ratio.}
\label{optmir}
\end{figure*}

\section{Bimodality in the infrared colours}
\label{sec:bimodal}

Given that for most galaxies the 24$\mu$m emission can be directly related to its current star-formation rate \citep[e.g.][]{calzetti07,kennicutt09,rieke09}, while its $K$-band flux can be considered to be a proxy for stellar mass \citep[e.g.][]{belldejong}, we might expect the $f_{24}/f_{K}$ infrared flux ratio to be a useful measure of the specific star-formation rate, and thus allow us to robustly separate passively-evolving and star-forming galaxies. 

In the left panel of Fig.~\ref{optmir} we plot the observed 24$\mu$m fluxes against the $K$-band magnitudes of spectroscopically confirmed supercluster members (green points), and those identified photometrically as from their UV-optical colours as possible cluster members (blue points). Following \citet{cortese08} who found that Coma cluster members could be efficiently distinguished from most background galaxies in the $g{-}i$ versus ${\rm NUV}{-}i$ observed colour-colour plot, we used the $B{-}R$ versus ${\rm NUV}{-}R$ observed colour-colour plot to produce an analogous separation between supercluster members and the background population. These colour cuts were then used to identify the photometric supercluster members, and by measuring the relative fractions of spectroscopically-confirmed supercluster and background galaxies satisfying these colour cuts, estimate the probability that each photometric member belongs to the supercluster (for details see Paper II).   

Along the top axis, we show the derived stellar masses, assuming a single $K$-band mass-to-light ratio of 0.5 \citep{belldejong}.
 The right-hand axis shows the corresponding SFRs derived using the infrared component of the combined FUV+24$\mu$m calibration of \citet{leroy}:
\begin{align}
{\rm SFR (M}_{\odot}{\rm yr}^{-1}) & = & 0.68{\times}10^{-28}L_{\nu}{\rm (FUV)[erg\,s}^{-1}{\rm Hz}^{-1}]\notag \\
& + & 2.14_{-0.49}^{+0.82}{\times}10^{-43}L(24\mu {\rm m)[erg\,s}^{-1}],
\label{leroy}
\end{align}
which assumes a Kroupa IMF. The first term comes from \citet{salim07}, while 
the obscured component of this calibration is essentially identical to that obtained by \citet{rieke09} of $2.04{\times}10^{-43}L(24\mu{\rm m)\,erg\,s}^{-1}$, and produces SFRs consistent within 10--20\% of those obtained via the $L_{TIR}$-based calibrations of \citet{bell03} and \citet{buat08}. 
The range of observed SFRs is then from 0.05\,M$_{\odot}\,{\rm yr}^{-1}$ at our 24$\mu$m survey limit of 350$\mu$Jy, up to 13\,M$_{\odot}\,{\rm yr}^{-1}$ for our brightest 24$\mu$m sources.

There appear to be two diagonal sequences (i.e. constant $f_{24}/f_{K}$) in the left panel of Fig.~\ref{optmir}: one at high stellar masses ($\mathcal{M}{\sim}10^{11}M_{\odot}$) and low 24$\mu$m fluxes (0.3--3\,mJy); and a second which contains all of the brightest 24$\mu$m sources, and which extends down to our survey limits at both 24$\mu$m and $K$-band wavebands. 
To elucidate better this bimodality we plot the infrared colour $f_{24}/f_{K}$ versus the 24$\mu$m flux for the same galaxies in the right-hand panel of Fig.~\ref{optmir}. Along the right-hand axis we present the distribution of $f_{24}/f_{K}$ flux-ratios of the 24$\mu$m-detected supercluster galaxies (in which the photometrically-selected galaxies are weighted according to the probability that they belong to the supercluster). The bimodal distribution in $f_{24}/f_{K}$ is clear, with two peaks at $f_{24}/f_{K}{\simeq}$1--6 and $f_{24}/f_{K}{\simeq}0$.03--0.15, and a clear gap at $f_{24}/f_{K}{\simeq}0$.2--0.4, where very few galaxies are located. Note that all $K{<}12$ supercluster galaxies are detected at 24$\mu$m (except two which lie in the wings of a nearby 24$\mu$m-bright source), so that the gap in the far lower-left corner of the left panel appears due to a real cut-off at $f_{24}/f_{K}{\sim}0.02$.

Given the well known bimodality in galaxy colours \citep{strateva,balogh04}, morphologies \citep{driver} and spectroscopic properties \citep{haines07} in which galaxies can be broadly split into blue, star-forming spirals and red, passively-evolving ellipticals \citep{blanton}, it would be natural to ascribe the observed bimodality in infrared colour as another manifestation of the red sequence / blue cloud division. We will discuss this later in \S~\ref{discbimodal}.

\begin{figure}
\centerline{\includegraphics[width=80mm]{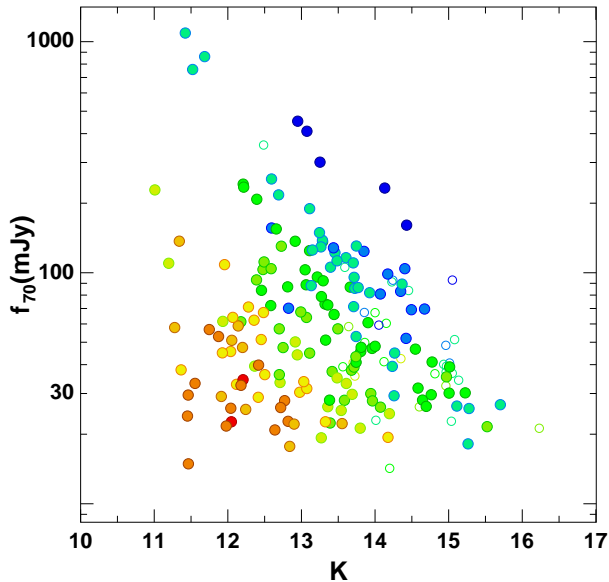}}
\caption{Observed 70$\mu$m fluxes versus $K$-band magnitude.  
Solid (open) symbols indicate spectroscopic (photometric) supercluster members. Each symbol is coloured according to its $f_{24}/f_{K}$ flux-ratio from red $({<}0.1)$ to blue $({>}8)$ as shown in Fig.~\ref{optmir}.}
\label{optmir70}
\end{figure}

In Figure~\ref{optmir70} we present the analogous plot to the left-hand panel of Fig.~\ref{optmir} for the 70$\mu$m far-infrared band, with spectroscopic (photometric) supercluster galaxies indicated by solid (open) symbols. Here as in subsequent plots, we colour each symbol according to its $f_{24}/f_{K}$ flux-ratio from red $({<}0.1)$ to blue $({>}8)$ as shown by the coloured vertical bar in Fig.~\ref{optmir}. 
Those galaxies which made up the star-forming sequence in Fig.~\ref{optmir} (green/blue symbols), dominate the 70$\mu$m cluster population, particularly at the bright end ($f_{70}{>}100\,$mJy), the sequence now extending over the full range of observed FIR fluxes (25--1000\,mJy). In terms of stellar mass, we are not detecting such low-mass galaxies as at 24$\mu$m, but follow the star-forming sequence over a dynamic range of ${\sim}40$ in stellar mass, extending down to $K{\sim}14$.5--15 (${\sim}K^{*}{+}3$).

At fainter FIR fluxes ($f_{70}{\la}100$\,mJy) we see an ever increasing contribution of high-mass ($K{\la}13$) ``transition'' galaxies with intermediate $f_{24}/f_{K}$ colours (yellow/orange symbols). We do not observe however, the same bimodality in galaxy properties as in Fig.~\ref{optmir}, as those galaxies forming the 24$\mu$m red-sequence remain undetected at 70$\mu$m.
This is not surprising given that the red sequence population only extends to $f_{24}{\sim}2.5$\,mJy, while \citet{temi09} find, for their sample of local elliptical galaxies, typical MIR colours of $(L_{24}[{\rm ergs/s}]/L_{70}){\sim}0$.2--2 or $(f_{70}/f_{24}){\sim}0$.6--6, which would suggest maximal 70$\mu$m fluxes of ${\sim}20$\,mJy, i.e. below our survey detection limit.

\subsection{Dusty star-formation on the optical red sequence}

The earliest studies of galaxy evolution within cluster environments often used optical colour as a proxy for star-formation history, identifying galaxies along the observed tight colour-magnitude (C-M) relation (red sequence) as passively-evolving, and those significantly bluer than the red sequence as star-forming \citep[e.g.][]{bower92,strateva,balogh04}, with the apparent increase in the fraction of blue galaxies in clusters out to $z{\sim}0.4$ used as evidence of a rapid evolution of the cluster population \citep[e.g.][]{bo78,bo84}. However, more recent studies have shown that a significant fraction (${\sim}3$0 per cent) of galaxies along the red sequence are actively forming stars, and appear red due to the presence of dust \citep{wolf05,haines08}. These dusty red sequence galaxies appear particularly prevalent in intermediate-density environments on the outskirts of clusters \citep{gallazzi,haines10}.

\begin{figure}
\centerline{\includegraphics[width=80mm]{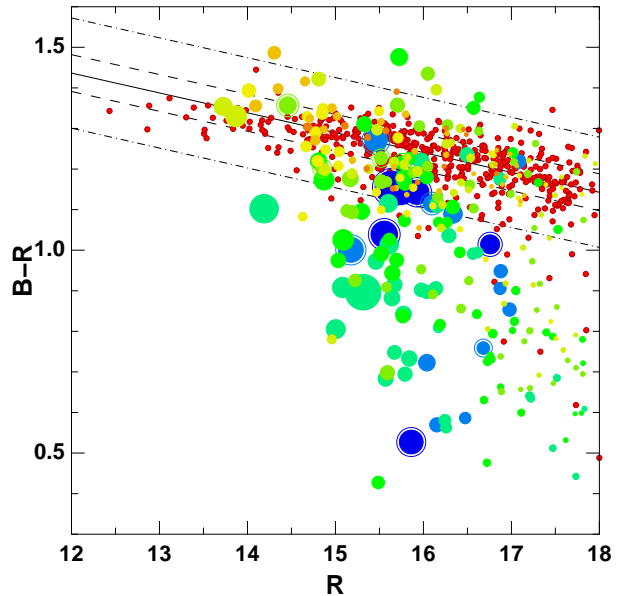}}
\caption{Optical $B{-}R/R$ colour-magnitude diagram for spectroscopic members of the Shapley supercluster. Symbols are coloured according to their $f_{24}/f_{K}$ ratio from red (${<}0.15$; including those not detected at 24$\mu$m) to blue (${>}8$) as shown in Fig.~\ref{optmir}. The size of the symbols (red ones excepted) scale with the 24$\mu$m flux. Ringed symbols indicate galaxies with $f_{24}{>}5$\,mJy whose 24$\mu$m remains unresolved by MIPS. The solid line indicates the best-fit CM-relation from \citet{sos2}, with the dashed and dot-dashed lines indicating the 1-$\sigma$ and 3-$\sigma$ dispersion levels respectively.}
\label{opticalcm}
\end{figure}

Figure~\ref{opticalcm} shows the optical $B{-}R/R$ C-M relation for spectroscopic supercluster members, the symbols coloured according to the $f_{24}/f_{K}$ as previously. The optical C-M relation fitted by \citet{sos2} (solid line) is closely followed by those galaxies identified by their infrared colours ($f_{24}/f_{K}{<}0.15$; red symbols) as passive, with very little scatter (${\la}0.05$\,mag), as expected. However, we also find that the transitional galaxies with $0.15{\le}f_{24}/f_{K}{<}1.0$ (yellow/orange symbols) also lie along the red sequence (and not below), albeit with somewhat greater scatter about the relation. While we see that the blue cloud population can be fairly associated to the star-forming sequence ($f_{24}/f_{K}{\ga}1$; green/blue symbols), there is also significant contamination of the red sequence by these actively star-forming galaxies \citep[c.f.][]{wolf05,haines08,mr09}.

In \citet{sos2} we separated the red sequence and blue cloud population using the line 3-$\sigma$ below the best-fit C-M relation (lower dot-dashed line in Fig.~\ref{opticalcm}). We find that 43 per cent of the obscured and 13 per cent of the unobscured star formation would be missed if we considered only the blue cloud population as star-forming. Similarly to \citet{wolf05,wolf} and \citet{haines08} we find that $30{\pm}4$ per cent of galaxies along the red sequence ($R{<}17$; above the 3-$\sigma$ lower limit) are identified as star forming ($f_{24}/f_{K}{>}0.15$), while $11{\pm}2$ per cent lie along or above the star-forming sequence ($f_{24}/f_{K}{>}1.0$). We note that this classification of red sequence galaxies only applies in the optical. If we were to examine the UV-optical C-M relation, these objects would now be classed as ``green valley'' or even blue cloud galaxies \citep{wyder07}.

\begin{figure*}
\centerline{\includegraphics[width=140mm]{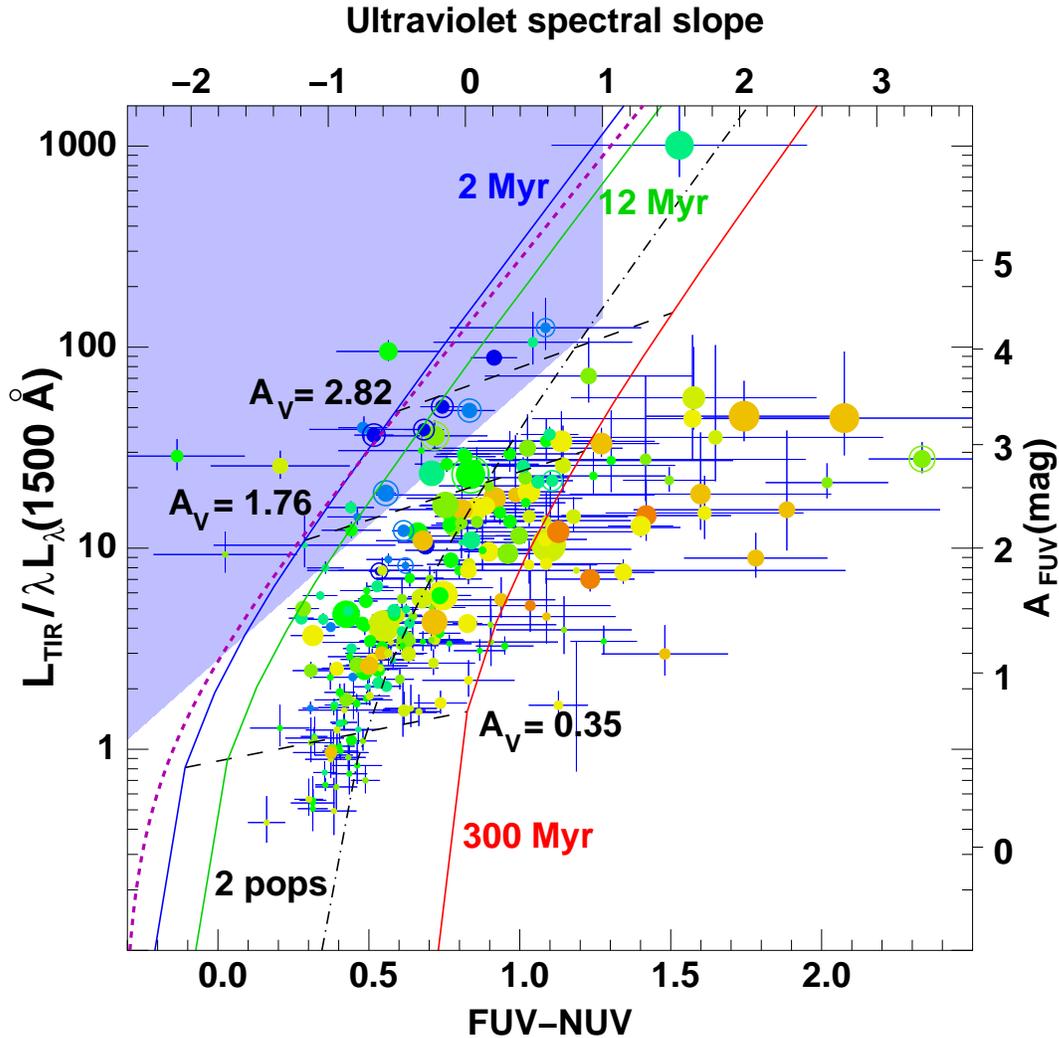}}
\caption{The IRX-$\beta$ plot for galaxies in the Shapley supercluster. Symbols are colour coded as in Fig.~\ref{optmir70} and are of sizes which scale with their $K$-band luminosity. Galaxies with $f_{24}{>}5$\,mJy for which the 24$\mu$m emission remains unresolved are ringed.
The magenta dashed line indicates the \citet{kong} starburst curve. The blue shaded region indicates the selection criteria used to identify our starburst galaxies. 
The solid lines indicate models of ageing instantaneous burst populations convolved with the starburst opacity curve for increasing amounts of dust attenuation from Fig. 10 of \citet{calzetti05} of 2 (blue line), 12 (green line) and 300 Myr old (red line). The dot-dashed line marked ``2pops'' shows the model track for the combination of a 5\,Myr old instantaneous burst with a 300\,Myr old one. The mass of the 5\,Myr old burst is 300 times lower than that of the 300\,Myr old one, and its extinction is systematically higher by $\Delta A_{V}=0.24$ mag; both population models are convolved with the starburst opacity curve. The far-ultraviolet extinction prescription used for the right-hand axis comes from \citet{buat}.}
\label{irxbeta}
\end{figure*}

\section{The IRX-$\beta$ relation}
\label{sec:irx}

A frequent difficulty in interpreting the observed UV-optical properties of galaxies comes from the difficulty in characterising the absorption of ultraviolet emission from young stars by dust. One issue relates to the wavelength-dependence of the attenuation curve, which determines how the same level of dust opacity is transformed into dust reddening, and does not appear to be universal. A second issue is that populations of different ages suffer different amounts of extinction, as young stars migrate with time from the highly obscured locations within giant molecular clouds, while the molecular clouds themselves begin to disperse with time, slowly unveiling their recently formed stars. A significant tool to understanding these issues has been the IRX-$\beta$ plot, which relates the overall level of dust opacity as measured by the infrared-to-ultraviolet ratio (IRX) to the observed amount of reddening by dust, as measured by the ultraviolet spectral slope, $\beta$. Here $\beta$ is defined by a power-law fit of the form $f_{\lambda}{\propto}\lambda^{\beta}$ to the ultraviolet spectrum at wavelengths $1300{\la}\lambda{\la}2600${\AA}. For an unobscured ongoing starburst, $F_{\lambda}{\propto}\lambda^{-2}$ in the ultraviolet \citep{kennicutt}, giving $\beta{\sim}{-}2$.

\citet{meurer} demonstrated that for local starburst galaxies there is a strong, tight correlation between their FIR-to-UV flux ratios and their ultraviolet spectral slopes, in which they redden systematically as more UV light is absorbed by dust. By producing an empirical fit to their IRX-$\beta$ curve, they suggest a simple calibration to estimate the level of dust extinction required to produce a given UV colour, allowing SFRs to be estimated from ultraviolet data alone. \citet{charlot} then showed that the IRX-$\beta$ relation can be understood most simply as a sequence in the overall dust content of galaxies. More recently however, it has been shown that quiescently star-forming galaxies (i.e. normal spirals) follow a different dust opacity-reddening relation than starburst galaxies, whereby their infrared-to-ultraviolet ratios are on average lower than those of starburst galaxies for the same ultraviolet colour, and also show a greater spread \citep{bell,kong,buat,dale07,boquien}. The ultraviolet colour however is quite sensitive also to the effective age of the stellar population, leading \citet{calzetti05} to suggest that whereas in starbursts, the ultraviolet emission comes predominately from current star formation processes, in normal star-forming systems it comes primarily from the more evolved, non-ionizing stellar population (${\sim}5$0--100\,Myr). 

In Figure~\ref{irxbeta} we show the IRX-$\beta$ plot for star-forming ($f_{24}/f_{K}{>}0.15)$ galaxies in the Shapley supercluster, the symbols colour coded according to their $f_{24}/f_{K}$ flux ratio as before, with sizes which scale with the $K$-band luminosity. The right-hand axis shows the resultant level of attenuation in the FUV estimated from the IRX based on the prescription described in Eq.~2 from \citet{buat}, which assumes that the bulk of dust emission is related to the ultraviolet absorption, as is the case for most normal star-forming galaxies. 
The bulk of our cluster galaxies that lie along the star-forming sequence of Fig.~\ref{optmir} (shown as green symbols) appear consistent with a single IRX-$\beta$ relation. This relation is however significantly offset from the empirical starburst relation of \citet{kong} indicated by the dashed magenta curve, having systematically redder ultraviolet colours by ${\sim}0.4$\,mag in $FUV{-}NUV$ for the same infrared-to-ultraviolet ratio. 
 The vast majority of our star-forming sequence cluster galaxies have their IR/FUV ratios and UV colours well described by the instantaneous burst populations of \citet{calzetti05} in the age range 12--300\,Myr (green and red solid lines), for a wide range of dust opacities, provided their dust attenuation and reddening characteristics are similar to those of the starburst galaxies. 

Given that both the 24$\mu$m and H$\alpha$ emission in normal star-forming galaxies are primarily produced in the H{\sc ii} regions around young ${<}10$\,Myr old O--B stars, this implies different stellar populations than those which dominate the UV stellar continuum emission. This can be partially reconciled by considering multiple stellar populations with a range of ages and dust extinctions, such as the two-population model of \citet{calzetti05} shown by the black dot-dashed curve. This consists of a 5\,Myr old burst component combined with a 300\,Myr old component in a mass ratio of 1:300. We see that the bulk of the galaxies along our star-forming sequence are also consistent with this two-population model, over a wide range of dust opacities, and show similar IRX-$\beta$ relations to other surveys of normal spiral galaxies \citep{calzetti05,dale07,johnson}. Galaxies with intermediate infrared colours lying between our star-forming and quiescent sequences (yellow/orange symbols) appear systematically redder in the ultraviolet than those on the star-forming sequence (green symbols) for a given IRX. This is consistent with them being more dominated by older stellar populations based on previous empirical correlations between the offset of a galaxy from the starburst IRX-$\beta$ relation and measures of star-formation history such as the 4000{\AA} break \citep{kong,johnson}.

Finally we note that ${\sim}1$0--20 (5--10\%) star-forming galaxies could be classed as starbursts, lying closer to the starburst curve of \citet{kong} than our nominal IRX-$\beta$ relation (indicated by the blue shaded region). Many of these have high $f_{24}/f_{K}$ values (indicated by their blue symbols), consistent with their having higher specific-SFRs than average, and with their classification as starburst galaxies. 

Although the resolution of the 24$\mu$m MIPS images is just 6\,arcsec, at least for the more luminous galaxies ($f_{24}{\ga}5$\,mJy) we can separate those galaxies resolved by MIPS and those for which the galaxies appear as point-sources, using the $\chi^{2}$ goodness-of-fit value produced by MOPEX when fitting the 24$\mu$m emission by the MIPS point spread function (see Paper II for details). This provides an independent way of distinguishing between extended emission from quiescent star formation in spiral disks and the unresolved emission produced by nuclear starbursts. Those galaxies with point-like 24$\mu$m emission are indicated as ringed symbols in Fig.~\ref{irxbeta}. 
The 24$\mu$m emission of many of the starburst galaxies in Fig.~\ref{irxbeta} is unresolved by MIPS and vice versa, suggestive of nuclear starbursts, unlike the remainder of the normal star-forming galaxies, for which the 24$\mu$m emission appears extended over the full galaxy disk.

In Paper II we estimated a total cluster SFR of $327\,{\rm M}_{\odot}\,{\rm yr}^{-1}$, of which $264\,{\rm M}_{\odot}\,{\rm yr}^{-1}$ (${\sim}80$ per cent) is obscured (based on the 24$\mu$m calibration) and just $63\,{\rm M}_{\odot}\,{\rm yr}^{-1}$ (${\sim}20$ per cent) is unobscured and emitted in the form of ultraviolet continuum. We estimate the contribution from the starburst galaxies (as identified above) to be 54\,M$_{\odot}\,{\rm yr}^{-1}$ or ${\sim}15$ per cent of the global star-formation within the cluster. 

\begin{figure*}
\centerline{\includegraphics[width=170mm]{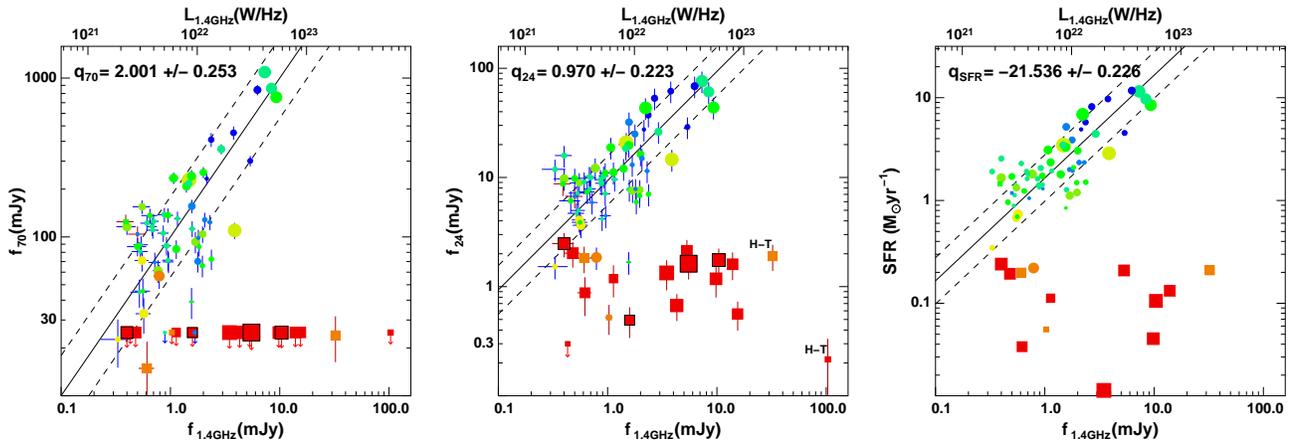}}
\caption{The FIR--radio correlation for known Shapley supercluster galaxies. 
({\em left}) The 70$\mu$m--1.4\,GHz FIR-radio correlation. ({\em centre}) The 24$\mu$m--1.4\,GHz FIR--radio correlation. ({\em right}) The UV+24$\mu$m-derived SFR-radio correlation.
Square symbols with red error bars indicate elliptical galaxies while circles with blue error bars indicate S0 or spiral morphologies. Symbols with black borders indicate BCGs. Head-tail radio galaxies are indicated as H-T.
The symbols are coloured according to their $f_{24}/f_{K}$ ratio as indicated in Fig.~\ref{optmir}, while their sizes scale with their $K$-band luminosity. Galaxies not detected in the mid-infrared are shown as upper-limits. The solid and dashed diagonal lines indicate the best-fit values of the FIR--radio  and SFR--radio correlations $q_{70}$, $q_{24}$ and $q_{SFR}$ and the $1\sigma$ spread, excluding those galaxies for which the radio emission is due to AGN.}
\label{radio}
\end{figure*}

\section{The FIR--radio correlation}
\label{sec:radio}

The radio population of galaxies is composed of a mix of sources powered by AGN and those whose radio emission traces its origin to star formation. 
Comprehensive surveys combining GHz radio continuum and {\em IRAS} far-infrared fluxes have found a tight linear relation between the observed radio  and far-infrared luminosities extending over five orders of magnitudes \citep{condon}, and including ${\ge}9$8\% of local field galaxies detected in the far-infrared \citep{yun}. This remarkably linear relation is generally believed to be driven by physical processes associated with young stars \citep[e.g.][]{condon91,dejong85,ivison10}.

Thermal (free-free) radio emission in galaxies is generally identified with the most massive short-lived stars \citep{condon}, but its contribution to radio luminosities at GHz frequencies is negligible. The non-thermal far-IR and radio emission share a common origin: massive and luminous stars and their products, namely supernova remnants, dust and cosmic rays. The young massive stars provide the UV radiation to heat the optically thick dust, which re-radiates in the FIR. The origin of non-thermal GHz radio continuum emission is synchrotron radiation from relativistic cosmic-ray electrons and protons, accelerated in supernova explosions of the same massive stars.  The correlation between the radio and FIR emission would, in this scenario, then require the timescale of starbursts to be suitably long (Gyrs).

In a detailed analysis, \citet{lacki10} show that the fact that relativistic particles and UV photons lose all of their energy before escaping these galaxies ("calorimetry") can be shown to be the underlying cause of the far-IR/radio correlation. However, as \citet{bell03} shows, for the relation to be valid over such a range of SFRs, a conspiracy between two other factors is required. In normal galaxies, where the average gas surface density $\Sigma_g$ is low, a significant fraction of relativitic electrons escape, leading to lower synchrotron radio emission, but this is compensated by lower opacity to UV photons, resulting in lower IR flux. At the high SFR end, where $\Sigma_g$ is high, bremsstrahlung and ionisation losses in the radio flux are compensated by emission from secondary electrons and protons.

In the cores of groups and clusters, this delicate balance can be disrupted due to the effect of the intergalactic medium on the $\Sigma_g$ of galaxies.  In many poor groups, where the relative velocities of galaxies are low, ram-pressure stripping is not important, and the radio and far-IR emission correlates with each other and with soft X-ray emission \citep[e.g.][]{sq09}, but the tidal effects of rich clusters and stripping processes affect these correlations.

\citet{niklas} examined the FIR--radio correlation of 45 galaxies in the Virgo cluster using {\em IRAS} FIR photometry in conjunction with radio measurements at 2.8 and 6.3\,cm, finding that most of the galaxies in the Virgo cluster obey the same FIR--radio correlation as field galaxies, and also have similar radio spectral indices. They also identified a few galaxies located in the inner part of the cluster with apparent excess radio emission, which they suggested could be due to ram-pressure compressing the magnetic field leading to enhanced synchrotron emissivity and/or reducing the amount of dust, resulting in a diminution of the FIR emission. \citet{mo01} examined the FIR--radio correlation for 329 radio galaxies from 18 nearby clusters, finding it to hold well for star-forming galaxies. They also found a rare but significant excess of star-forming galaxies in the cluster centres for which their radio emission was enhanced typically by 2--3${\times}$, suggestive again of the magnetic field strengths of these galaxies being increased by compression from ram-pressure. 

We take our radio data from the published catalogue of 1.4\,GHz radio continuum sources identified by \citet{miller} from his survey of the Shapley supercluster core with the Very Large Array (VLA). The VLA survey covers a 7\,deg$^{2}$ region which encompasses all of our UV--FIR datasets, and reaching 5$\sigma$ sensitivities of 330$\mu$Jy at a resolution of 16\,arcsec. Of these 79 are supercluster members detected at 24$\mu$m.

Figure~\ref{radio} shows the 1.4\,GHz radio continuum flux plotted versus the 70$\mu$m ({\em left panel}) and 24$\mu$m ({\em middle panel}) far-infrared fluxes for Shapley supercluster galaxies, colour coded according to their $f_{24}/f_{K}$ flux ratios as in Fig.~\ref{optmir}. Square symbols indicate galaxies morphologically classified as ellipticals, while circles indicate spiral or S0 morphologies. Galaxies not detected in the infrared are shown as upper-limits. It is immediately apparent that the radio active galaxies constitute two quite different populations.

The first radio population is infrared dim, with $f_{24}{\la}2$mJy and $f_{70}{\la}25$mJy ($L_{IR}{\la}5{\times}10^{9}L_{\odot}$). These 18 galaxies have low $f_{24}/f_{K}$ ratios placing them along the red sequence of passive galaxies identified in Fig.~\ref{optmir}, for which we expect the infra-red emission to be due to dust heated by evolved stars rather than current star-formation. These are all massive galaxies ($\mathcal{M}{\ga}10^{11}{\rm M}_{\odot}$), morphologically classified as early-type, including four cD galaxies. Given the morphologies, and absence of mid-infrared emission, we assume that the radio emission in this population comes from AGN activity.

The second radio population comprises 61 infrared-bright spirals, being detected at both 24$\mu$m and 70$\mu$m wavelengths, and show clear correlations between the infrared and radio continuum fluxes. The solid line in each panel shows the best fit linear correlation between the FIR and radio fluxes, of the form $\log f_{FIR}({\rm Jy}){=}\log f_{1.4{\rm GHz}}({\rm Jy})+q_{FIR}$, after excluding the infrared-dim radio AGN population described above. We obtain best-fit $q$-parameters of $2.00{\pm}0.25$ at 70$\mu$m and $0.97{\pm}0.22$ at 24$\mu$m, the 1$\sigma$ scatter in the relation indicated by dashed lines in each panel. These values and scatters are consistent with those obtained by \citet{appleton} and \citet{marleau} of $q_{70}=2.15{\pm}0.16$ and $q_{24}=0.84{\pm}0.28$ based on a sample of more than 500 matched IR/radio sources with redshifts from the {\em Spitzer} extragalactic First Look Survey. The scatter in the FIR--radio correlation is also consistent with the 0.26\,dex obtained by \citet{yun} for their FIR-selected ({\rm IRAS} $f_{60{\mu}m}{>}2$\,Jy) complete sample of 1809 galaxies with redshifts and 1.4\,GHz radio fluxes. 

In the right panel of Fig.~\ref{radio}, we compare the 1.4\,GHz radio fluxes of the supercluster galaxies directly with their SFRs obtained from the combined FUV and 24$\mu$m data via the calibration of \citet{leroy}, after correcting the 24$\mu$m fluxes for emission from dusty evolved stars assuming the linear correlation between $f_{24}$ and $f_{K}$ obtained by \citet{temi09} for passive elliptical galaxies ($f_{24}{\sim}0.03f_{K}$). Given that for almost all these galaxies, the infrared contribution to the SFR is larger than the ultraviolet one, this plot is almost unchanged from the central panel comparing $f_{24}$ with $f_{1.4{\rm GHz}}$. We obtain a best-fit SFR calibration for the radio continuum of:
\begin{equation}
{\rm SFR}_{\rm 1.4\,GHz} (M_{\odot}{\rm yr}^{-1}) = \frac{L_{1.4{\rm GHz}}}{3.44{\times}10^{21}{\rm W\,Hz}^{-1}},
\end{equation}
and an rms scatter of 0.226\,dex about the relation. The denominator is midway between those obtained by \citet{hopkins} and \citet{rieke09} ($2.88{\times}10^{21}$ and $3.84{\times}10^{21}$ respectively) after correcting the latter two by a factor 1.59 to account for the different IMF used.
Based on this, the 330$\mu$Jy 5$\sigma$ limit of the 1.4\,GHz survey corresponds to a star formation rate of ${\sim}0.5\,{\rm M}_{\odot}\,{\rm yr}^{-1}$, an order of magnitude less sensitive than either the {\em GALEX} FUV or {\em Spitzer} 24$\mu$m datasets.

\begin{figure}
\centerline{\includegraphics[width=80mm]{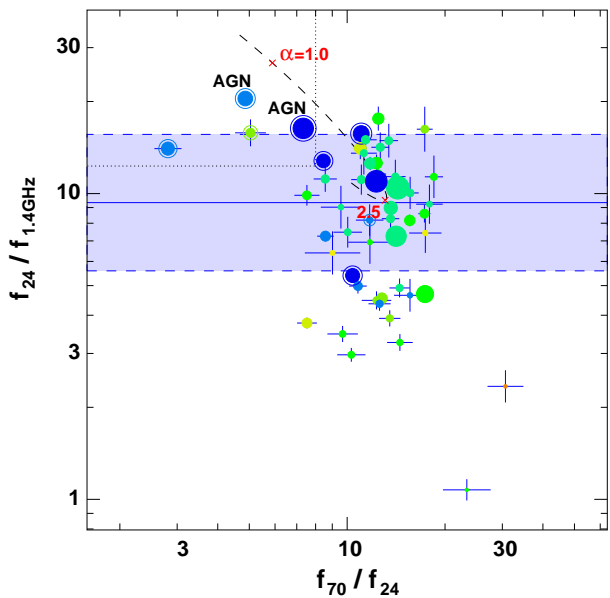}}
 \caption{Relation between $q_{24}$ and the mid-infrared colour $f_{70}/f_{24}$ of supercluster galaxies. The size of each symbol scales with $f_{24}$, while the colour indicates the $f_{24}/f_{K}$ flux ratio as before, with sources unresolved at 24$\mu$m ringed. The two galaxies optically classified as AGN from the spectra of \citet{smith07} are indicated.
The blue line and shaded region indicate the obtained 24$\mu$m FIR--radio relation and 1-$\sigma$ scatter. The dashed curve indicates the expected FIR--radio colours for the \citet{dh02} model SEDs, with the colours corresponding to the $\alpha{=}1$ (AGN-heated dust) and 2.5 (quiescent star-formation) models indicated.}
\label{radioircols} 
\end{figure}

The obtained FIR--radio correlation indicates that SFRs can be estimated for cluster star-forming galaxies from their monochromatic 24$\mu$m, 70$\mu$m infrared or 1.4\,GHz radio fluxes, to an accuracy of the order 0.24\,dex. However, given the significant population of radio-loud AGN, which dominate at the highest radio luminosities, it will be vital to be able to distinguish between AGN and star-formation as the power source for the radio emission. For regions lacking infrared data, the best approach would be to use optical morphology to distinguish the populations, identifying AGN with early-types and star-forming galaxies with spirals. 
\cite{yun} suggest using a limit of $L_{1.4\,GHz}{=}10^{23}\,{\rm W\,Hz}^{-1}$ to statistically divide the lower-luminosity starburst-spiral population from the higher-luminosity AGN population, while \cite{morrison} suggest that for clusters where galaxies with high SFRs are rare, a slightly lower dividing limit of $L_{1.4\,GHz}{=}10^{22.75}\,{\rm W\,Hz}^{-1}$ is more suitable. However, as can be seen from Fig.~\ref{radio}, while we do not find star forming spirals with radio luminosities above these limits, and there remain a significant population of AGN down to the survey limit of $L_{1.4\,GHz}{=}10^{21.3}\,{\rm W\,Hz}^{-1}$.
It would be dangerous to use a blind radio survey to examine the effects of environmental processes on cluster galaxies, without being able to separate the star-forming and AGN sub-populations. For example, in \citet{morrison}'s radio survey of 30 Abell clusters with $z{\le}0.25$, Abell 1689 is identified as being more radio active than the sample average at the 90\% level, but nine out of the ten confirmed radio-loud cluster galaxies are identified as being elliptical galaxies powered by AGN rather than star-forming spirals. 

If one or more of the 24$\mu$m, 70$\mu$m far-infrared or 1.4\,GHz radio fluxes have significant contributions from dust heated by evolved stars or AGN in a major subset of the population, then we may expect to see systematic trends when plotting their flux ratios against one another. In this case, the SFR estimates based on one or more of these fluxes may contain some systematic biases. To investigate this, we plot in Fig.~\ref{radioircols}, the quantity $q_{24}$ versus the far-infrared colour $f_{70}/f_{24}$ for those supercluster galaxies detected at all three wavelengths. For the vast majority (${\sim}90$ per cent) of the population, there is no apparent correlation between $q_{24}$ and far-infrared colour, which suggests that the dominant contributor to all three fluxes for these galaxies is star formation. Their colours are reasonably consistent with the FIR emission being produced by dust heated by the relatively low-intensity radiation fields found in galaxies quiescently forming stars across the full extent of their disks, as described by the $\alpha{=}2.5$ model SED of \citet{dh02}.

There are also four galaxies with systematically hotter $f_{70}/f_{24}$ colours and excess 24$\mu$m above expectations from the FIR--radio correlation (indicated by the region delineated by the dotted line). These colours ($f_{70}/f_{24}{\la}8$), correspond to the {\em IRAS}-based mid-infrared excess criteria ($S_{25\mu{\rm m}}/S_{60\mu{\rm m}}{\ge}0.18$) empirically proposed as in indicator of an infrared AGN \citep{degrijp}, and the flux ratios are most consistent with the $\alpha{=}1$ model SED of \citet{dh02} in which the FIR emission comes from dust heated by an intense radiation field of an AGN. These sources are all unresolved in the 24$\mu$m images (as indicated by ringed symbols), and the two for which we have AAOmega spectra from \citet{smith07} are both classified as AGN from their location in the \citet{bpt} diagnostic diagram.

\section{Discussion}\label{sec:discuss}

\subsection{The bimodality in galaxy infrared colours}
\label{discbimodal}

In Figure~\ref{optmir} we showed that the infrared colours of the supercluster galaxies to be bimodally distributed in $f_{24}/f_{K}$, with two peaks well separated by a clear gap where few galaxies are located. 
Given the well known bimodality in galaxy colours \citep{strateva,balogh04,wyder07,haines08}, morphologies \citep{driver} and spectroscopic properties \citep{haines07} in which galaxies can be broadly split into blue, star-forming spirals and red, passively-evolving ellipticals \citep{blanton}, it would be natural to ascribe the observed bimodality in infrared colour, which can be considered a proxy for specific-SFR, as another manifestation of this division. 
In this case, we would expect the $f_{24}/f_{K}$ flux ratio to correlate strongly with morphology or other indicators of star-formation history such as UV-optical colours or spectral indices including EW(H$\alpha$). 

\subsubsection{Correlation with morphology}

\begin{figure}
\centerline{\includegraphics[width=80mm]{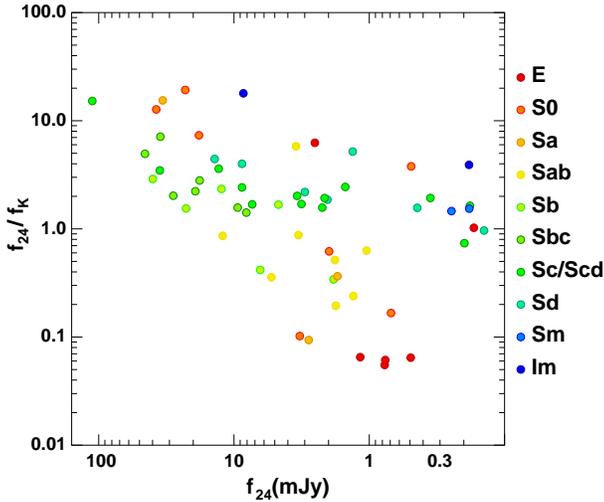}}
\caption{The infrared colour $f_{24}/f_{K}$ versus $f_{24}$ flux of galaxies from the SINGS sample, where $f_{24}$ flux is based upon the galaxy being shifted to the distance of the Shapley supercluster. The colour of the symbol indicates its morphological class from red (E) to blue (Im), as indicated.}
\label{sings}
\end{figure}

One approach to better understand what Fig.~\ref{optmir} is telling us, is to plot it for a sample of local galaxies whose properties are well understood and defined. For this purpose, we take the {\em Spitzer} Infrared Nearby Galaxies Survey (SINGS) of 75 nearby ($<30$\,Mpc) galaxies \citep{sings}, for which comprehensive spatially-resolved infrared and complementary observations are available \citep{dale07}. Fig.~\ref{sings} replots the $f_{24}/f_{K}$ versus $f_{24}$ diagram of Fig.\ref{optmir} for the SINGS sample, adjusting the infrared fluxes out to the distance of the Shapley supercluster, and colour codes the symbols according to the galaxy's morphological classification. Although the actual bimodality is not quite reproduced, we can now identify the sequence at $f_{24}/f_{K}{\sim}0.05$ with the passive E/S0 population, while the sequence at $f_{24}/f_{K}{\sim}2$ corresponds broadly to the mass sequence of Sb--Sd spirals and irregulars, albeit along with a number of earlier-type spirals and dusty S0s. Intermediate between the two main sequences we find primarily  Sab spirals.

\citet{li07} examined the correlations between the infrared colour, ${\nu}L_{\nu}(24\mu{\rm m})/{\nu}L_{\nu}(3.6\mu{\rm m})$ (i.e. similar to our $f_{24}/f_{K}$ flux ratio) and morphology for a limited sample of 154 galaxies with both SDSS-DR4 spectroscopy and mid-infrared photometry from SWIRE. They observed a similar bimodality in the infrared colour to that seen in Fig.~\ref{optmir}, which robustly separates galaxies according to their morphology with ${\sim}8$5\% reliability, either visually classified as early- (E/S0) and late-types (Sa--d and Irr) or according to their bulge-to-total ratio.

\subsubsection{Correlation with $FUV{-}R$ colour and EW(H$\alpha$)}

\begin{figure}
\centerline{\includegraphics[width=80mm]{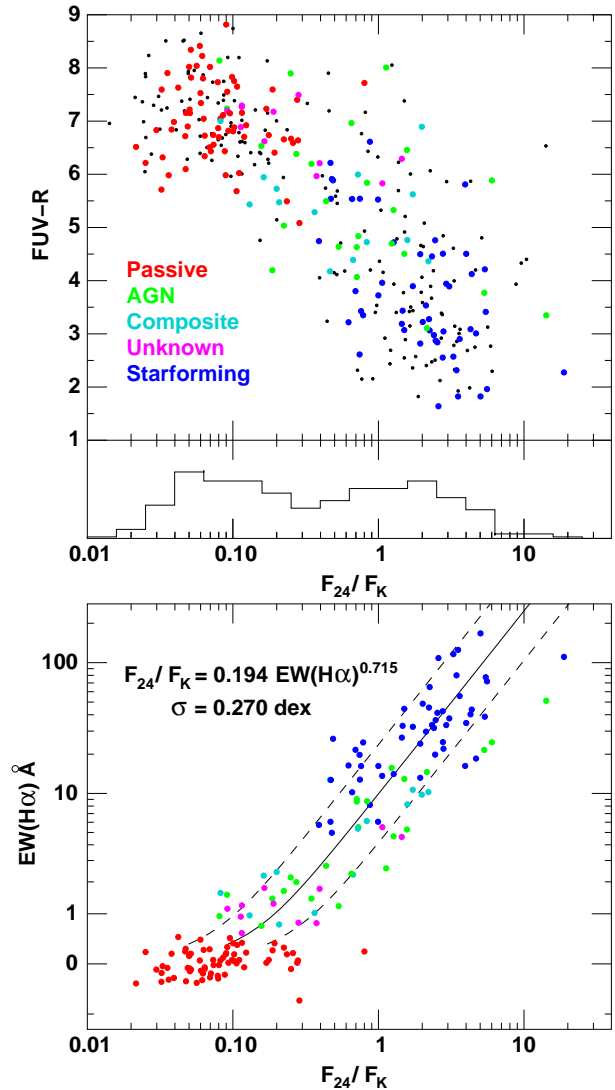}}
\caption{Correlations between the $f_{24}/f_{K}$ MIR--$K$-band flux ratio with $FUV{-}R$ UV-optical colour ({\em top panel}) and EW(H$\alpha$) obtained from the AAOmega spectroscopy ({\em bottom panel}). The colour of each symbol indicates the corresponding spectroscopic classification: ({\em red}) passive; {\em green} AGN; ({\em cyan}) composite; ({\em magenta}) unknown; ({\em blue}) star-forming, while black points in the top panel are SSC members not observed with AAOmega.}
\label{ha_mir}
\end{figure}

\begin{figure*}
\centerline{\includegraphics[width=170mm]{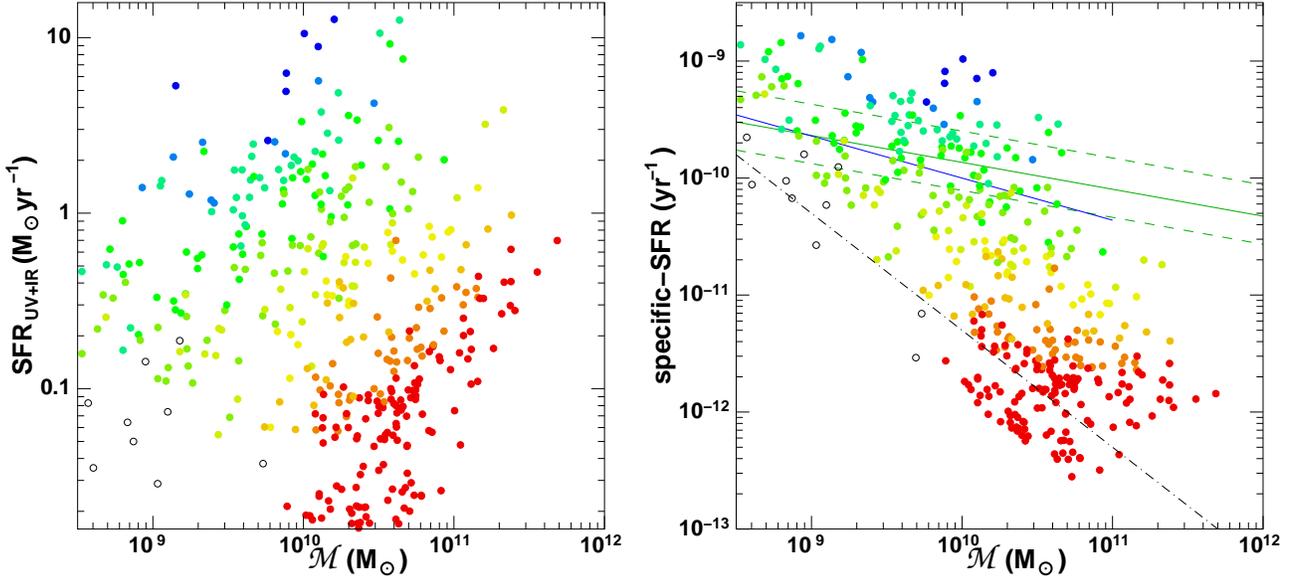}}
\caption{Relationship between SFR ({\em left panel}) and specific-SFR ({\em right panel}) with stellar mass for the supercluster galaxies. The SFR is the sum of the SFR(FUV) and SFR(24$\mu$m) components when galaxies are detected at 24$\mu$m (filled symbols) and SFR(FUV) only otherwise (open symbols). Stellar masses are derived from the $BRK$ photometry as described in \citet{merluzzi}. The filled symbols are coloured according to $f_{24}/f_{K}$ as in Fig.~\ref{optmir}. The black dot-dashed line indicates the SFR corresponding to our 24$\mu$m completeness limit. The green solid and dashed-lines indicate the SSFR--$\mathcal{M}$ relation and scatter of local ($0.015{<}z{\le}0.1$) blue star-forming SDSS galaxies of \citet{elbaz}, while the blue line indicates the corresponding relation of \citet{schiminovich}.}
\label{ssfr}
\end{figure*}

We further examine the nature of the bimodality in Fig.~\ref{ha_mir}, which compares $f_{24}/f_{K}$ with the $FUV{-}R$ UV-optical colours ({\em top}) and H$\alpha$ equivalent widths\footnote{Following \citet{haines07} the y-axis is scaled as  sinh$^{-1}$\,EW(H$\alpha$): this results in the scale being linear at EW(H$\alpha){\sim}0${\AA} where measurement errors dominate; and logarithmic at EW(H$\alpha){\ga}5${\AA}.} ({\em bottom}) obtained from the AAOmega spectroscopy of \citet{smith07}, using the emission-line ratios to classify galaxies following the diagnostic diagrams of \citet{bpt}. 
Considering simply the ranges of $f_{24}/f_{K}$ for the diverse spectroscopic classifications, it is immediately apparent that we can robustly split galaxies into star-forming or passive populations about a flux ratio $f_{24}/f_{K}{\sim}0.3$. We can thus identify the two sequences seen in Fig.~\ref{optmir} as: (i) the blue cloud of star-forming galaxies with $f_{24}/f_{K}{\sim}0$.5--10; and (ii) the red sequence of passively-evolving galaxies having $f_{24}/f_{K}{\sim}0$.02--0.3. This is confirmed in the top panel, which shows the strong correlation between $FUV{-}R$ colour and $f_{24}/f_{K}$, whereby galaxies on the infrared passive sequence ($f_{24}/f_{K}{<}0.15$) all lie on the UV--optical red sequence ($FUV{-}R{\simeq}7$), whereas galaxies on the infrared star-forming sequence ($f_{24}/f_{K}{\sim}2$), lie within the UV--optical blue cloud ($FUV{-}R{\la}5$). We can specifically exclude a population of actively star-forming galaxies with low-dust contents that would place them on our infrared passive sequence ($f_{24}/f_{K}{<}0.15$), by the lack of objects in the lower-left corner. Additionally, we note that there are no SSC galaxies which are not-detected at 24$\mu$m, but are FUV-bright, having $FUV{-}R{<}5$.

We can quantify the correlation between $f_{24}/f_{K}$ and EW(H$\alpha$) which, if we exclude the passive galaxies (for which there is no significant H$\alpha$ emission), results in a tight power-law (solid line in Fig.~\ref{ha_mir}) of the form:
\begin{equation}
f_{24}/f_{K}{=}0.194{\rm EW(H}\alpha)^{0.715}
\end{equation}
with a spread of $\sigma=0.27$\,dex in $f_{24}/f_{K}$ at fixed EW(H$\alpha$) (indicated by the dashed-lines). This power-law correlation appears valid over a factor ${\ga}100$ in EW(H$\alpha$), and holds for galaxies classified as either star-forming or AGN, although AGN appear to have marginally higher $f_{24}/f_{K}$ ratios than star-forming galaxies of comparable EW(H$\alpha$), perhaps suggesting that the H$\alpha$ emission is more obscured in AGN. The observed deviation from a linear relation between ($f_{24}/f_{K}$) and EW(H$\alpha$) would suggest that galaxies with lower values of EW(H$\alpha$) are more obscured (assuming that the 24$\mu$m emission comes from star formation), or that less of the 24$\mu$m emission is due to star formation (which is likely the case for the passive galaxies). 

\subsubsection{Comparison to the specific-SFR--$\mathcal{M}$ relation of star-forming local field galaxies}

In Figure~\ref{ssfr} we express the observed bimodality in $f_{24}/f_{K}$ in terms of specific-SFRs, deriving SFRs from the FUV and 24$\mu$m fluxes using Eq.~\ref{leroy} and taking the stellar masses from \citet{merluzzi}, plotting the SFR--$\mathcal{M}$ ({\em left panel}) and specific-SFR--$\mathcal{M}$ ({\em right panel}) relations. Unsurprisingly, given that our SFRs and stellar masses are derived respectively from the 24$\mu$m and $K$-band luminosities, the $f_{24}/f_{K}$ flux ratio (indicated by the colour of each symbol) correlates strongly with specific-SFR, allowing us to relate each value of $f_{24}/f_{K}$ to an associated value or range in the current to past-averaged SFR (birthrate parameter) for that galaxy. Moreover, it allows us to directly compare the properties of our observed cluster star-forming and passive sequences, with their equivalents derived for local field galaxies using diverse multi-wavelength datasets and approaches to estimate SFRs and stellar masses \citep[e.g][]{brinchmann,elbaz,salim07,schiminovich,bothwell}.

 Although not as distinct as in Fig.~\ref{optmir} the observed cluster star-forming sequence (green/blue symbols in the right panel of Fig.~\ref{ssfr}) appears consistent with the local field specific-SFR--$\mathcal{M}$ relations of \citet{elbaz} (solid and dashed green lines) and \citet{schiminovich} (blue line), with specific-SFRs in the range 0.3--1$0\times10^{-10}\,{\rm M}_{\odot}{\rm yr}^{-1}/{\rm M}_{\odot}$, while the red sequence galaxies (red symbols) have ${\rm SSFRs}{\le}3{\times}10^{-12}\,{\rm M}_{\odot}{\rm yr}^{-1}/{\rm M}_{\odot}$. \citet{vulcani} compare the specific-SFR--$\mathcal{M}$ relations between cluster and field environments at $0.4{<}z{<}0.8$, and although their results appear strongly affected by their 24$\mu$m completeness limits impinging on the star-forming sequence, the two relations do appear consistent.

The clear gap separating the two sequences of Fig.~\ref{optmir} has mostly disappeared. We believe this is due mostly to systematic uncertainties due to dust extinction on the stellar mass estimates obtained from our $BRK$ photometry.
 However, given the robust separation in the {\em observable} $f_{24}/f_{K}$ ratio, it is likely that with a stellar-mass estimation approach better able to account for the effects of dust obscuration, this gap would be reproduced in the {\em derived} specific-SFR distribution.

Looking again at the observed bimodality in $f_{24}/f_{K}$ of Figure~\ref{optmir}, it seems that the gap between the two sequences is clearest over $12.5{\la}K{\la}15$, corresponding to intermediate masses with $\mathcal{M}{\sim}10^{10}{\rm M}_{\odot}$, whereas at higher masses ($\mathcal{M}{\sim}10^{11}{\rm M}_{\odot}$) the bimodality could be said to disappear, with a significant population of ``transition'' galaxies with intermediate $f_{24}/f_{K}$ colours at $K{\sim}12$. The paucity of intermediate-mass transition objects suggests that the time-scale for quenching star-formation in these galaxies is rather brief, so that they pass rapidly from the star-forming to passive sequence. The increasingly populated transition region at high masses instead suggests that here quenching time-scale is much longer, such that many more galaxies are observed {\em during} the transformation process.  It is not clear however if this is due to environmental processes related to galaxies encountering the cluster environment, as these trends bear resemblance to those seen for local field galaxies. \citet{lee} and \citet{bothwell} find that the distribution of specific-SFRs for galaxies on the star-forming sequence is narrowest ($\sigma_{SSFR}{\sim}0.4$\,dex) at intermediate masses ($8.5{\la}{\log}{\mathcal M}({\rm M}_{\odot}){\la}10$) and luminosities ($-15{>}M_{B}{>}{-}19$). Above the characteristic transition stellar mass of ${\sim}3{\times}10^{10}{\rm M}_{\odot}$ where the population is dominated by early-type spirals, there appears a turnover to lower specific-SFRs, similar to those seen in our red sequence population. \citet{bothwell} show that this is a consequence of the much reduced H{\sc i} gas content and shorter gas consumption time-scales of high mass galaxies, suggesting that the transition to low-specific SFRs is due to the innate earlier and more rapid conversion of gas into stars of massive galaxies \citep[see also][]{haines07}. Further observations (such as those being undertaken within ACCESS) will be required to determine if our ``transition'' massive galaxies are there due to internal processes related to their high masses (and which would occur in any environment), or instead due to current interaction with the supercluster environment quenching their star formation now.

\subsubsection{The MIR red sequence population}
\label{redsequence}

If we na\"{i}vely assume the 24$\mu$m emission is due to star-formation, it would seem surprising we find that all of the massive ($\mathcal{M}{\sim}10^{11}M_{\odot}$) early-type galaxies have significant detections at 24$\mu$m, given that they show no H$\alpha$ emission, and even more surprising that they show a sequence of constant $f_{24}/f_{K}$ or specific-SFRs. 

One plausible important contribution to the 24$\mu$m emission is simply the mid-infrared photospheric emission from evolved stars that would be expected based on the Rayleigh-Jeans extrapolation of a black body at 3--5000K. In this case expected infrared flux-ratios due solely to photospheric emission is found to be $f_{24}/f_{K}=0.014$ \citep{helou04,draine07}, a factor of a few lower than that observed for the red sequence population, and hence unlikely to be the dominant contributor to the 24$\mu$m emission for the vast majority of these galaxies. 
However {\em Spitzer} MIPS and Infrared Spectrograph observations of early-type galaxies have shown that the diffuse, excess emission apparent at 10--30$\mu$m in these galaxies is due to silicate emission from the dusty circumstellar envelopes of mass-losing evolved asymptotic giant branch stars \citep{bressan,temi,clemens}, rather than residual ongoing star-formation. The strength of this silicate emission is a slowly declining function of stellar age \citep{piovan03}, and persists even for the very evolved stellar populations (${>}1$0\,Gyr) identified from optical spectroscopy, resulting in infrared colours consistent with those found Virgo and Coma clusters \citep{clemens}, as well as those observed here. 
It is thus likely that the observed SFRs in the range 0.1--1\,M$_{\odot}{\rm yr}^{-1}$ and specific-SFRs in the range 0.3--3${\times}10^{-12}\,{\rm M}_{\odot}{\rm yr}^{-1}/{\rm M}_{\odot}$ of the observed cluster red-sequence galaxies are significantly overestimating the actual values, due to the contribution of evolved stars.

\subsection{The nature of star formation in clusters}

One of the key aims of this and other infrared surveys of galaxies in group/cluster environments (e.g. XI groups, LoCuSS) is to obtain complete censuses of star-formation in dense environments, and to establish whether there is a population hidden from previous UV/optical surveys with significant obscured SFRs sufficiently large to build up an S0 bulge in the time-scale required to transform the star-forming spirals seen in $z{\sim}0.4$ clusters into the passive S0s that have empirically replaced them in local clusters \citep[e.g.][]{dressler97,geach06,geach09,smith10}. In the massive cluster Cl\,0024+16 at $z{\sim}0.4$ \citet{geach09} identified a population of heavily obscured LIRGs with 24$\mu$m-derived SFRs of 30--60\,M$_{\odot}\,{\rm yr}^{-1}$ and {\em Spitzer}/IRS mid-infrared spectra consistent with nuclear starbursts. Via IRAM CO interferometric observations of two of these LIRGs, they found significant cold molecular gas, sufficient to build up an S0 bulge in as little as 150\,Myr. Going out to $z{\sim}1$ even higher SFRs appear frequent in cluster galaxies \citep{marcillac,koyama}.

We find no such galaxies with SFRs in the range 20--40\,M$_{\odot}\,{\rm yr}^{-1}$ (converting to our Kroupa IMF) within the Shapley supercluster, comparable to those found by \citet{geach09}. All the supercluster members are observed to have SFRs (including both obscured and unobscured components) in the range 0.1--13\,M$_{\odot}\,{\rm yr}^{-1}$ (Figs.~\ref{radio} and~\ref{ssfr}), and specific-SFRs implying mass-doubling time-scales ${\ga}1$\,Gyr. Given that we are spectroscopically complete for all sources with $f_{24}{>}26$\,mJy (irrespective of UV--optical colour or morphology), we can be confident that we are not missing any supercluster galaxy within our mid-infrared survey area having a ${\rm SFR}{\ga}4$\,M$_{\odot}\,{\rm yr}^{-1}$, while the observed FIR--radio relation (Fig.~\ref{radio}) indicates consistency of our SFRs with those derived from the radio. 

This limit to relatively modest SFRs (${\la}15$\,M$_{\odot}\,{\rm yr}^{-1}$) has been found for galaxies in all local clusters analysed to date in the mid-infrared out to $z{\sim}0.2$, including Coma \citep{bai06}, Abell 3266 \citep{bai}, Abell 901/2 at $z{=}0.17$ \citep{wolf} and Abell 1689 at $z{=}0.18$ \citep{duc,haines10}. Of the eight $z{\sim}0.2$ clusters of \citet{smith10} for which robust SFRs have been derived combining H$\alpha$ and 24$\mu$m fluxes for a highly complete spectroscopic sample, just one galaxy has been found with SFRs similar to those found by \citet{geach09}, the BCG in Abell 1835 whose ${\rm SFR}{\sim}70\,{\rm M}_{\odot}\,{\rm yr}^{-1}$ \citep{pereira} is due to its specific situation of being fed by the massive cooling flow. This wholesale reduction in the SFRs of cluster galaxies over this period simply reflects the ${\sim}10{\times}$ cosmic decline in SFRs since $z{\sim}1$ \citep{zheng,lefloch}, and in particular the ${\sim}10{\times}$ decrease in the density of LIRGs and ULIRGs over the last 4 Gyr \citep{lefloch}. This means that if S0 bulges are to have been built up by star formation since $z{\sim}0.4$, this could have occurred rapidly over a few ${\times}10^{8}$yr at $z{\sim}0.4$ or if taking place later ($z{\la}0.2$) must be taking place over much longer time-scales (${\ga}1$\,Gyr).

A number of mechanisms have been proposed to transform infalling spiral galaxies into passive S0s which include intermediate starburst and possible subsequent post-starburst phases \citep[see e.g.][]{miller}. For example, starbursts could be initiated in galaxies during the early phases of ram-pressure stripping as the interaction of the intracluster gas act to compresses their molecular clouds \citep{dressler83}, while gravitational shocks induced by repeated high-velocity encounters \citep[harassment;][]{moore} or low-velocity encounters / mergers \citep[often in infalling groups --- pre-processing;][]{balogh04b} could induce a bar driving gas inwards and feeding a nuclear starburst. 

Observationally \citet{moss} indicated that 50--70 per cent of the infalling population in clusters were interacting or merging and triggering star formation.  \citet{fadda} and \citet{porter,porter08} found enhanced star formation along filaments feeding rich clusters and suggested that these dense galaxy environments favoured gravitational interactions that could induce the observed starbursts.
\citet{moran} identified a population of starburst galaxies at the cluster virial radius, and suggested that these were triggered by interaction between the infalling galaxy and the shock boundary between the hot, dense ICM and the colder, diffuse infalling gas from outside the cluster.  
\citet{roettiger} predicted that when clusters merge they develop shocks in the ICM, which could lead to starbursts in galaxies as they pass through them, a theory supported by a number of MIR/radio-based surveys who found an excess of dusty star-forming galaxies in merging clusters \citep{mo01,miller,haines09b,johnston}, but see also \citet{haines09} and \citet{rawle} for counterexamples..

We identified starburst galaxies in the Shapley supercluster core via three methods: (i) their location above the specific-SFR--$\mathcal{M}$ relation or equivalently having $f_{24}/f_{K}{\ga}5$ (blue symbols in Figs.~\ref{optmir70} and~\ref{irxbeta}); (ii) their location along the starbust IRX--$\beta$ relation of \citet{kong} (blue shaded region of Fig.~\ref{irxbeta}); or (iii) having unresolved 24$\mu$m emission indicative of a nuclear starburst (indicated by ringed symbols in Fig.~\ref{irxbeta}). There is significant overlap among these three groups indicating good consistency between these classifications: two-thirds of the $f_{24}/f_{K}{\ga}5$ galaxies above $f_{24}{=}5$mJy (where we can separate resolved and point-like sources) remain unresolved; while 60\% of galaxies consistent with the starburst IRX--$\beta$ also have $f_{24}/f_{K}{\ga}5$. Independent of which method we consider, we identify ${\sim}15$ starburst galaxies across the survey, contributing just ${\sim}1$5\% of the global SFR within the supercluster core. We defer an investigation of the triggering of these starbursts to a future work, but note here that their location primarily along the filament connecting A3562 and A3558 is indicative of a environment-related cause.

The dominant contributor (${\sim}8$5 per cent) to the global SFR budget in the Shapley supercluster is rather spiral galaxies undergoing normal, quiescent star formation across their disks, in a manner indistinguishable from the general field population. This is manifest firstly by the star-forming sequence identified in Figure~\ref{optmir}, and which comprises essentially all the star-formation in the supercluster, being indistinguishable from that obtained from the sample of local field spiral galaxies observed within SINGS (green symbols in Fig.~\ref{sings}), while the derived specific-SFRs (Fig.~\ref{ssfr}) are consistent with the specific-SFR--$\mathcal{M}$ relations of local field galaxies obtained by \citet{elbaz}, \citet{schiminovich} and \citet{bothwell}. Secondly the mid-infrared colours ($f_{70}/f_{24}{\sim}15$; Fig.~\ref{radioircols}) and FIR--radio ratios (Fig.~\ref{radioircols}) observed for most galaxies on the star-forming sequence are best described by the $\alpha{\sim}2.5$ model infrared SEDs of \citet{dh02}, corresponding to dust heated by relatively low-intensity and diffuse star-formation taking place over an extended disk \citep{dale}, as observed for typical local spirals in the SINGS sample \citep{dale07}. Finally, in the IRX--$\beta$ plot of Figure~\ref{irxbeta} we find that the bulk of the star-forming galaxies follow a single dust opacity-reddening relation that is however significantly offset from the empirical starburst relation of \citet{kong}, having systematically redder ultraviolet colours for the same level of obscuration (IRX). The observed IRX--$\beta$ relation is similar to those seen by other surveys of local spiral galaxies, and is consistent with a contribution to the UV flux from a dominant older stellar population. 

These three independent lines of evidence from diverse parts of the spectral energy distribution all confirm that the bulk of star-forming galaxies in the supercluster are forming stars in a manner indistinguishable from the general field population, consistent with them being galaxies recently accreted from the field, but yet to encounter the dense ICM or be affected by cluster-related processes.
 This is consistent with the conclusions drawn by \citet{haines09} from the observed composite radial gradients in the fraction of MIR luminous star-forming galaxies ($f_{SF}$) for 30 massive clusters at $0{<}z{<}0.3$. The observed gradual increase of $f_{SF}$ with clustercentric radius was found to be consistent to first order with the predicted radial trends from numerical simulations in which the star-forming galaxies are identified with galaxies infalling into massive clusters for the first time. From these simulations, the fraction of infalling galaxies expected to be observed in cluster surveys is seen to increase approximately linearly with projected radius to around twice the virial radius \citep[to be compared with the trends of e.g.][]{lewis}. This association of the cluster star-forming population with the infalling field population was first made by \citet{balogh00} and \citet{ellingson} to explain the origin of the SF-density relation and the observed radial population gradients in star-formation rates and colours of cluster galaxies within the simple model of clusters being built up through the accretion of field galaxies. As we discussed in Paper II, the overall normalization of the infrared LF and hence cluster SFR is ${\sim}2$5--3$0{\times}$ that which we would expect if the star-forming galaxies were simply interlopers whose line-of-sight velocities happen to lie within the redshift range of the supercluster. Hence they must instead be physically associated with the observed overdensity. Equally they are not evenly distributed over the whole survey region, but preferentially lie along, or close to, the filamentary structure connecting the clusters Abell 3558 and Abell 3562, if not within the cores of either cluster. This points towards them being generally located in the infall regions, some $\sim$0.5--5\,Mpc from their nearest cluster \citep[c.f.][]{biviano}.

The domination of the observed cluster star-forming population by normal star-forming spiral galaxies who are on their first infall into the cluster and have yet to be affected by the dense environment makes it much more difficult to disentangle those galaxies which are {\em currently} in the process of being transformed from star-forming spiral to passive lenticular (perhaps via a short starburst phase), and ultimately distinguish among the possible physical mechanisms behind this transformation. In particular it limits the sensitivity of the infrared or ultraviolet luminosity functions to these processes, and could explain the full consistency between the infrared luminosity functions observed in local clusters \citep[Coma, Abell 3266 and Shapley;][Paper II]{bai06,bai} and those seen in the field, as discussed Paper II. However we identify a few candidate sub-populations which could represent diverse phases in this transformation: (i) galaxies with intermediate infrared colours ($0.15{\le}(f_{24}/f_{K}){<}1.0$) lying between the star-forming and passive sequences, covering a region in colour-space that a star-forming galaxy must pass through as it is being quenched; (ii) the nuclear starburst population, which could indicate a short phase of star-formation triggered via an interaction or bar instability which could serve to build up the bulge of an S0. These sub-populations are the prime candidates for our ongoing integral-field spectroscopic survey with WiFeS, and will be investigated further in future papers.

\section{Summary and prospects}

We have presented an analysis of panoramic {\em Spitzer}/MIPS mid- and far-infrared (MIR/FIR) and {\em GALEX} ultraviolet imaging of the most massive and dynamically active structure in the local Universe, the Shapley supercluster at $z{=}0.048$, covering the five clusters which make up the supercluster core. Combining this with existing spectroscopic data from 814 confirmed supercluster members and 1.4\,GHz radio continuum maps, these represent the largest complete census of star-formation (both obscured and unobscured) in local cluster galaxies to date, extending down to ${\rm SFRs}{\sim}0.$02--0.$05\,{\rm M}_{\odot}{\rm yr}^{-1}$. We have taken advantage of this comprehensive panchromatic dataset and census of star formation, in order to perform the kind of detailed analysis of the nature of star formation in cluster galaxies, previously limited to local field galaxy surveys such as SINGS \citep{dale07,draine07}.

We observed a robust bimodality in the infrared ($f_{24{\mu}{\rm m}}/f_{K}$) galaxy colours, which we were able to identify as another manifestation of the broad split into star-forming, blue spiral and passive, red elliptical galaxy populations seen in UV-optical surveys.
This diagnostic also allows the identification of galaxies in the process of having their star formation quenched as the infrared analogue of the ultraviolet ``green valley'' population. The bulk of supercluster galaxies on the star-forming sequence have specific-SFRs consistent with local field specific-SFR--$\mathcal{M}$ relations of Elbaz et al. (2007), and form a tight FIR--radio correlation (0.24\,dex) confirming that their FIR emission is due to star formation. 

Using several quite independent diagnostics of the quantity and intensity of star formation, we have developed a coherent picture of the types of star formation found within cluster galaxies, showing that the dominant mode, contributing 85 per cent of the global SFR, is quiescent star formation within spiral disks, as manifest by the observed sequence in the $(L_{IR}/F_{FUV}$)--($FUV{-}NUV$) plane -- the IRX-$\beta$ relation -- significantly offset from the starburst relation of \citet{kong}, while their FIR--radio colours indicate dust heated by low-intensity star formation. Just 15 per cent of the global SFR is due to nuclear starbursts, that could represent star formation triggered by interaction of galaxies with their environment. The vast majority of star formation seen in cluster galaxies comes from normal infalling spirals who have yet to be affected by the cluster environment.

In future works we will examine the environmental trends in star formation, and identify and quantify various classes of galaxies in the process of being transformed from star-forming spiral to passive S0. These transition galaxies are being targeted for integral-field spectroscopy as part of a long-term programme using the WiFeS IFU-spectrograph \citep{dopita07} on the 2.3-m ANU telescope at Siding Spring Observatory, Australia.
These detailed observations will allow the mechanism by which the global SFR in these galaxies is being quenched to be resolved and identified. For example, ongoing ram-pressure stripping can be seen as knots of H$\alpha$ emission being swept off the galaxies \citep{cortese07,sun,yoshida} by their passage through the dense ICM \citep{kronberger,kapferer,tonnesen}, resulting in truncated H$\alpha$ profiles \citep[e.g.][]{vogt}, beyond which the characteristic post-starburst signatures of deep Balmer absorption-lines may be found \citep{crowl}.

\section*{Acknowledgements}

This work was carried out in the framework of the collaboration of the FP7-PEOPLE-IRSES-2008 project ACCESS. CPH, RJS and GPS acknowledge financial support from STFC.  GPS acknowledges support from the Royal Society. We thank Michael Dopita for useful discussions during the writing of this work.
This work is based in part on observations made with the Spitzer Space Telescope, which is operated by the Jet Propulsion Laboratory, California Institute of Technology under a contract with NASA.
{\em GALEX (Galaxy Evolution Explorer)} is a NASA Small Explorer launched in April 2003. We gratefully acknowledge NASA's support for construction, operation and science analysis for the {\em GALEX} mission, developed in cooperation with the Centre National d'Etudes Spatiales of France and the Korean Ministry of Science and Technology.

\label{lastpage}
\end{document}